\documentclass{svproc}
\usepackage{url}

\usepackage{amssymb}

\usepackage{hyperref}

\usepackage{graphicx}        % standard LaTeX graphics tool
\def\Journal#1#2#3#4{{#1} {\bf #2}, #3 (#4)}

\def\JHEP{{J. High Energy Phys.}}
\def\PLB{{ Phys. Lett.}  B}
\def\PRL{ Phys. Rev. Lett.}
\def\PRD{{ Phys. Rev.} D}

\def\PR{{ Phys. Rep.}}

\def\NIMA{{ Nucl. Instrum. Methods Phys. Res., Sect. A }}

\def\qarko{q\bar{q}}

\begin{document}
\mainmatter
\title{Prospects in spectroscopy with Belle II}
%\titlerunning{Hamiltonian Mechanics}

\author{Vishal Bhardwaj \\ (for the Belle II Collaboration)}
\authorrunning{Vishal Bhardwaj}
%for an abbreviated version of
\institute{Indian Institute of Science Education and Research Mohali, Punjab 140306, India.\\ \email{vishstar@gmail.com} }
%\and Name of Second Author \at Name, Address of Institute \email{name@email.address}}
\maketitle
\begin{abstract}
Belle played a leading role in shaping the spectroscopy sector for last
decade. With 50 times more data than Belle, the Belle II experiment also
expects to play crucial role in spectroscopy for the next
decade. In this talk, a few chosen results one expects from Belle II
will be discussed.
\end{abstract}

%\abstract{Each chapter should be preceded by an abstract (10--15 lines long) that summarizes the content. The abstract will appear \textit{online} at \url{www.SpringerLink.com} and be available with unrestricted access. This allows unregistered users to read the abstract as a teaser for the complete chapter. As a general rule the abstracts will not appear in the printed version of your book unless it is the style of your particular book or that of the series to which your book belongs.\newline\indent
%Please use the 'starred' version of the new Springer \texttt{abstract} command for typesetting the text of the online abstracts (cf. source file of this chapter template \texttt{abstract}) and include them with the source files of your manuscript. Use the plain \texttt{abstract} command if the abstract is also to appear in the printed version of the book.}

\section{Introduction}

The Belle II detector~\cite{belle2_detector} is a general purpose detector built to
test Standard Model mechanism by doing precision measurements.  Belle II
also provides a very clean environment and  is an ideal place to carry
quarkonium $\qarko$ spectroscopy related studies. $\qarko$ are produced through $B$
decays, double charmonium production, two photon production, initial state radiation, and quarkonium decay/transitions.

For the last 15 years Belle~\cite{abashian} (predecessor of the Belle II detector with similar environment)
had a very successful program on quarkonium ($q\bar{q}$).
Many new $q\bar{q}$ (-like) states such as $\eta_c(2S)$, $X(3872)$,
$X(3915)$, $Z(3930)$, $X(3940)$,  $Z_1(4050)^+$, $Y(4260)$, $Z(4430)^+$,
$Y(4660)$, $Z_b(10610)$, and $Z_b(10650)$ have been found. Many of these
states cannot be accomodated by the conventional spectroscopy.
Some states have non-zero charge which suggest that they are
tetraquark/molecule-like states. Belle II (with the ability to accumulate 50 times more data in comparison
to Belle) will be able to play important role in understanding the nature of
these states. In this talk, I will try to give brief overview of the
Belle II program for quarkonium. I should admit here that I have not done
justice in this proceeding.
Interested readers should refer to the Belle II Physics book~\cite{B2Physics}.

\section{Belle to Belle II}

The Belle II experiment (situated in Tsukuba, Japan), is the upgraded successor
of Belle.  The detector's major upgrades in comparison to Belle are:
\begin{itemize}
\item {A Vertex detector (VXD) consisting of two layers of DEPFET pixels
  (PXD) and  four layers double-sided silicon strips (SVD), with improved
  resolution (to half) compared to Belle.}
\item{A central drift chamber (CDC) with larger volume drift chamber, smaller
  drift cells and faster electronics.}
\item{Completely new particle identification [time of propagation (barrel)
  and proximity-focusing Aerogel Ring-Imaging Cherenkov detector (end-cap)].}
\item{Belle CsI (Tl) crystals are used for the electro-magentic calorimeter with
  modified waveform sampling electronics to reject pile-up events.
}
\item{Upgraded $K_L-\mu$ detection system (KLM) where resistive plate counter
  used in Barrel. Because of the projected inefficiency of RPCs at high
  ambient rate, Belle II end-caps are instrumented with scintillator strips.}
  %~\cite{Arxiv1011.0325,S. Hashimoto, (ed. ) et al. , KEK-REPORT-2004-4}}
\end{itemize}

\section{Current status of Belle II}

Belle II successfully completed ``Phase II'' commissioning runs and
accumulated
472 pb$^{-1}$ of data. During Phase II, all the
sub-detectors were in except the full vertex detector
(partial vertex detector for a  particular $\phi$ was in).

\begin{figure}[h!]
  \includegraphics[height=40mm,width=60mm]{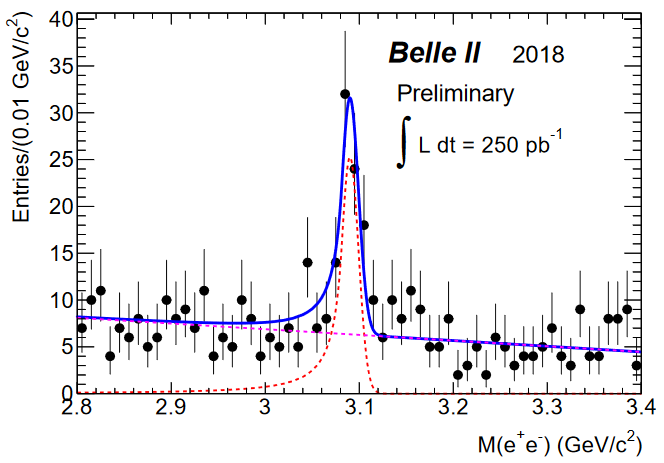}
  \includegraphics[height=40mm,width=60mm]{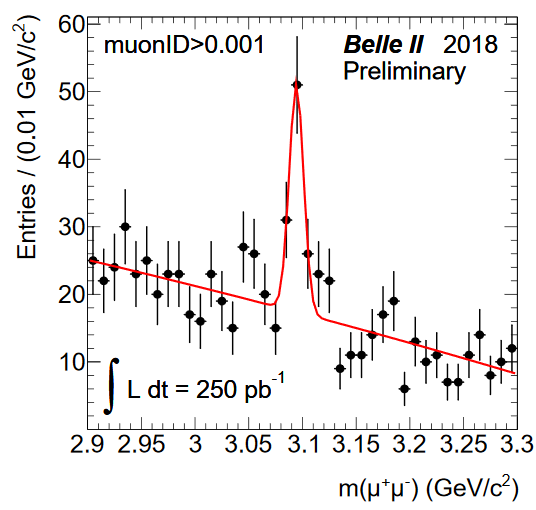}
  \caption{Reconstructed invariant mass of $J/\psi \to e^+e^-$ (left)
    and $J/\psi \to \mu^+ \mu^-$ (right) at Belle II using partial Phase II
    data. We also have the plots with full data at current date
    (however, the plots shown here are similar to what was shown at the
    conference). Plots with full data set can be found at
    Ref.~\cite{public_plots}}
  \label{fig:1}       % Give a unique label
\end{figure}

\subsection{Re-discovery of ``November revolution''}
Figure~\ref{fig:1} shows the reconstructed $J/\psi \to \ell^+ \ell^-$
demonstrating the capability of reconstructing lepton tracks. We see
a clear peak of $J/\psi$ to $e^+e^-$ and $\mu^+ \mu^-$ reconstruction.

\subsection{Re-discovery of $D$ and $B$ mesons}
\begin{figure}[h!]
  \includegraphics[height=40mm,width=60mm]{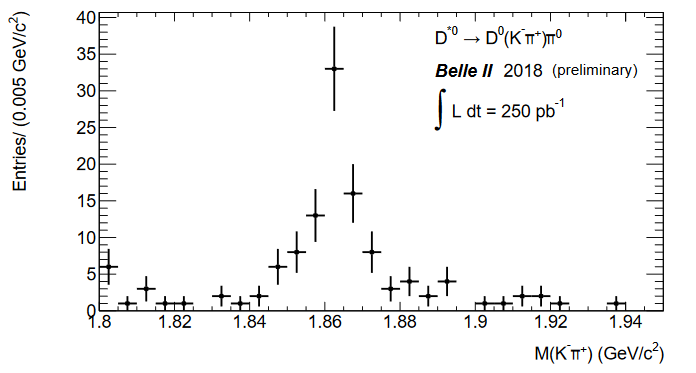}
  \includegraphics[height=40mm,width=60mm]{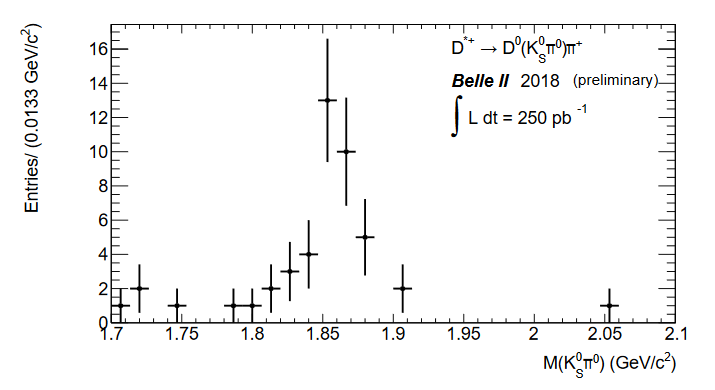}
  \includegraphics[height=40mm,width=60mm]{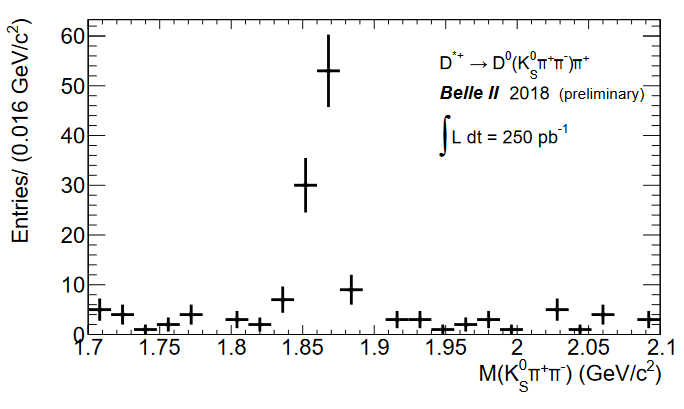}
  \includegraphics[height=40mm,width=60mm]{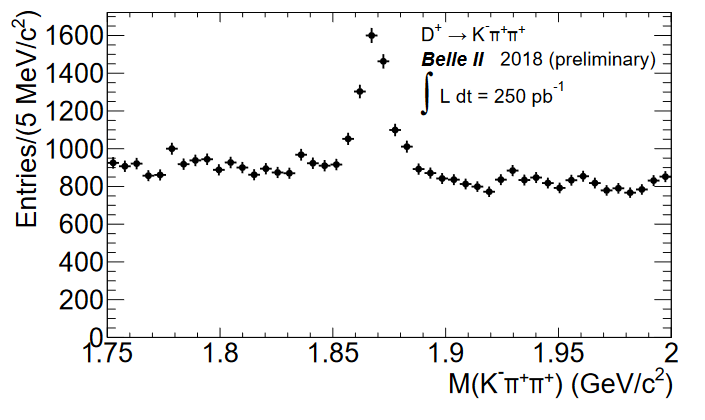} 
  \caption{Reconstructed invariant mass of $D$ mesons from various decay modes.}
  \label{fig:2}       % Give a unique label
\end{figure}

\begin{figure}[h!]
  \includegraphics[height=40mm,width=60mm]{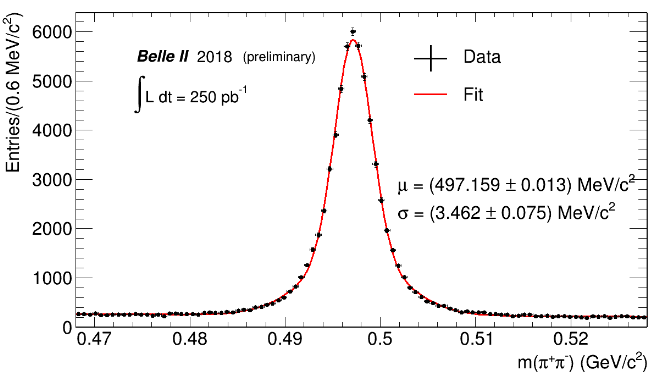}
  \includegraphics[height=40mm,width=60mm]{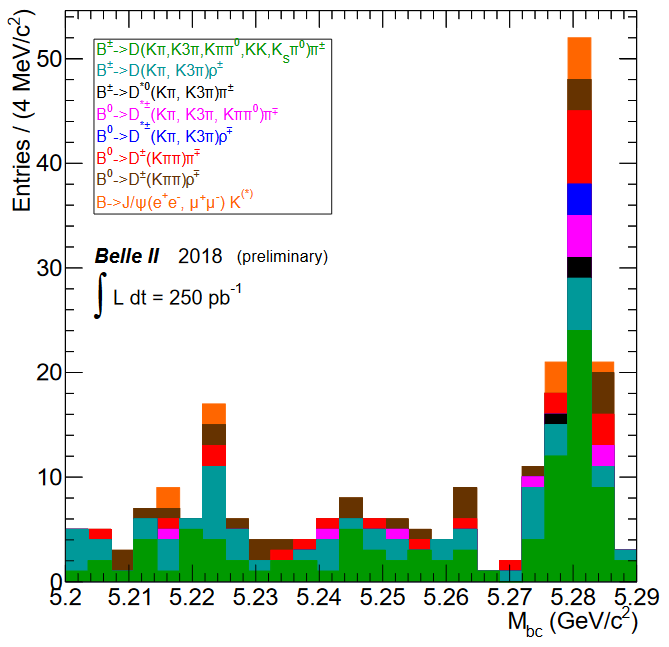}
  \caption{Invariant mass of reconstructed $K_S^0 \to \pi^+\pi^-$ (left) and
    $M_{\rm bc}$ for reconstructed $B$ meson from different
    modes.}
  \label{fig:3}       % Give a unique label
\end{figure}

Figure~\ref{fig:2}-\ref{fig:3} shows the reconstructed $D$ and $B$ mesons
demonstrating the capability of reconstructing charged and neutral Kaon and
pions.

As seen from the re-discovery plots of the $J/\psi$, $D$, and $B$, the Belle II
detector is working as per expectation.

\section{Prospects for $c\bar{c}$(-like) states}

$X(3872)$ was first observed in $B^+ \to (J/\psi \pi^+ \pi^-) K^+$ process at 
Belle~\cite{Choi_PhysRevLett.91.262001}. Soon after its discovery, 
$X(3872)$ was confirmed by CDF~\cite{Acosta_PRL.93.072001}, 
D$\varnothing$~\cite{do1},  BaBar~\cite{babar0}, LHCb~\cite{lhcb} and CDF~\cite{cms}.
A lot of effort went into studying this particle, thanks to which now we
know  its precise mass, width, and $J^{PC}$ to be 
$(3871.69\pm 0.17)$ MeV$/c^2$~\cite{PDG}, $< 1.2$ MeV~\cite{Choi_PRD84.052004},
and $1^{++}$~\cite{AAij_Jpc}, respectively.
At Belle II, we expect 1500 signal events with 10 ab$^{-1}$ of data (which is
1/5 of the total data Belle II aims to accumulate). Just to give an idea, the current
yield of $B^+ \to \psi'(\to J/\psi \pi \pi)K^+$ is ~3600 signal events at Belle.
This will help in measuring precisely $X(3872)$ mass and width.
Within the first two years of data taking, one can expect that Belle II will
accumulate 5 to 10 ab$^{-1}$ of data.

The narrow width of $X(3872)$ and the proximity of its mass to the 
$D^0 \bar{D^*}$ threshold makes it a good candidate for a
$D^0 \bar{D^*}$ molecule~\cite{swanson}. 
Currently the most probable explanation for the $X(3872)$ nature is a
molecule with
admixture of charmonium.

If  $X(3872)$ is charmonium then one expects it to be $\chi_{c1}'$. If
so then it should decay to $\chi_{c1} \pi^+ \pi^-$. Current search by the Belle
has a negative result~\cite{vishal_1}. One can measure or expect a
tighter constraint from Belle II.

Performing the study of $X(3872) \to \bar{D^0}D^{*0}$~\cite{aushev}
with the full Belle II
data will bring more information. Measuring the ratios of radiative
decays~\cite{radiative}
$\mathcal{B}(X(3872) \to \psi' \gamma)/\mathcal{B}(X(3872)\to J/\psi \gamma)$
with more data is what Belle II should do, as it is crucial for understanding
the nature of $X(3872)$. 
If $X(3872)$ is a $D^0 \bar{D}^{*0}$ molecule, then
one expects that there may be other  ``$X$-like'' particles with different
quantum numbers that are bound states of  $D^{(*)}$ mesons, such as
a $(D^0 \bar{D}^{*0}  - \bar{D}^0D^{*0})$ combination as the $C$-odd partner of
$X(3872)$ with $J^{PC}$ of $1^{+-}$. $C$-odd search has been negative till
now~\cite{negsearch}. Searching for the the charged $X(3872)\to J/\psi\pi^+\pi^0$~\cite{Choi_PRD84.052004} and $C$-odd
partners such as  $J\psi \eta$ at Belle II is interesting. If found, it will
suggest a molecular/tetraquark  nature of the $X(3872)$~\cite{Maiani}.
On the other side,
absence of charged partners suggest $X(3872)$ to be an iso-singlet state. This
suggests $X(3872) \to J/\psi \pi^+ \pi^-$ to be an iso-spin violating decay.
BaBar has measured the ratio
$\mathcal{B}(X(3872)\to J/\psi \omega(\to \pi^+\pi^-\pi^0))/\mathcal{B}(X(3872)\to J/\psi \pi^+\pi^-))$ $=0.8\pm0.3$. Belle II can improve this ratio with
much precision.

Absolute $\mathcal{B}(B\to X(3872)K^+)$ helps in measuring
$\mathcal{B}(X(3872)\to \textrm{final states})$. This measurement is only
possible at the $e^+e^-$ $B$ factories.  One has to reconstruct the missing
mass recoiling against the $K^+$,

\begin{equation}
  M_{\rm miss} = \sqrt{ (p^*_{e^+e^-} - p^*_{\rm tag} -p^*_K)^2}/c
\end{equation}

where $M_{\rm miss}$ is the missing mass recoiling against the $K^+$, and $p^*_{e^+e^-}$, $p^*_{\rm tag}$, and $p^*_K$ are the four-momenta of the electron-positron
initial state, $B_{\rm tag}$ (full reconstruction of one of the two
charged $B$ mesons via hadronic states) and kaon, respectively, in the center-of-mass frame. The $M_{\rm miss}$ peaks around the mass of the signal. Belle
measured $\mathcal{B}(B^+ \to X(3872) K^+)$ $< 2.6 \times 10^{-4}$ (@ 90\%
CL)~\cite{Kato}. With 50 times more data, Belle II can measure the branching
fraction till $10^{-5}$ or less due to the improvement~\cite{TKeck} in the full
reconstruction algorithm (in comparison to Belle).

Not only decays, but also production of $X(3872)$ in the $B$ decay
provide information about the nature of $X(3872)$. Belle observed the
$B^0 \to X(3872) K^+ \pi^-$ decay mode having $7\sigma$ significance.
In their
study of  the production dynamics of $B^0 \to X(3872) K^+ \pi^-$, they found
that $B^0 \to X(3872) K^*(892)^0$ does not dominate the
$B^0 \to X(3872) K^+ \pi^-$ decay, which is in contrast
to the normal charmonium states (where $K^*(892)^0$  dominates)~\cite{Anu}.
This suggest that $X(3872)$ doesn't behave like normal charmonium 
states. With 10 ab$^{-1}$ of data collected with Belle II, we expect
$B\to X(3872) K \pi$ to have the same number of events
to what Belle has accumulated for $B\to \psi' K \pi$.  Therefore,
one can expect to have a more precise measurement.

In the two photon process, $\gamma \gamma \to J/\psi \phi$,  Belle observed
$X(4350)$~\cite{Belle_2p}. However, recently in the amplitude analysis of
$B \to J/\psi \phi K$, LHCb found several structures
($Y(4140)$, $Y(4274)$, $X(4500)$, and $X(4700)$) but did not found $X(4350)$~\cite{LHCb_2p}.
Belle II should revisit with more data. Another area where Belle II can
contribute is the $Y(4260)$ study. Belle II will compliment BESIII here.
We expects improvement in mass resolution due to 
CDC improvements. Belle II with 50 ab$^{-1}$ should be able to study the line-shape of
$Y(4260)$. Another possible study one can think of is
$e^+e^- \to Y(4260)(\to J/\psi \pi^0\pi^0) \gamma_{ISR}$ to search for a neutral partner.
Also, measuring $\mathcal{B}(B \to Y(4260) K)$ at Belle II is an important in
step to understand the nature of $Y(4260)$. The first charged $Z(4430)^+$ state
was seen by Belle in  the $B^0 \to (\psi' \pi^+) K^-$ decay
mode~\cite{choi_PRL_100_142001}. 
Till recently this state was not well established due to non observation in
other experiments. Recently, LHCb confirmed $Z(4430)^+$ and using an Argand
diagram, they supported the resonance nature of this state~\cite{LHCb_Z4430}.
Belle II can perform amplitude analyses with more statistics (similar to the
one done at Belle~\cite{Chilikin_ZJP,Chilikin_JpiK}) and help in
understanding these states with precision. 
Other modes not feasible at Belle are also accessible at Belle II.
For example only with 10 ab$^{-1}$ of data at Belle II, one expects the
yield of the $B^0 \to (\chi_{c2} \pi^-) K^+$ decay mode to become comparable
to what Belle accumulated for
$B^0 \to (\chi_{c1} \pi^-) K^+$~\cite{Belle_chic1pi}. Not only that but
Belle II can also search for  the neutral partners using $\pi^0$ modes 
($B^0 \to (c\bar{c}) \pi^0 K^+$).

\section{Prospects for $b\bar{b}$(-like) states}

The bottomonium spectrum has found to be different from what we have
understood in charmonium spectrum. Belle II is a unique place to carry out
bottomonium related studies due to the energy accessible by SuperKEKB
(expecting to reach $\Upsilon(5,6S)$ energy). We know that $Z_b$ states were found
in the $\Upsilon(5S)$ decays by Belle and are clear signature of exotic states.
Belle~\cite{Adachi_1} found that 
\begin{equation}
  % \[
  \frac{\Gamma(\Upsilon(5S) \to h_b(nP) \pi^+ \pi^-)}{\Gamma(\Upsilon(5S) \to \Upsilon(2S) \pi^+ \pi^-)} = \left\{\begin{array}{lr} 0.45\pm 0.08^{+0.07}_{-0.12},  & \textrm{for }  h_b(1P) \\  0.77\pm 0.08^{+0.22}_{-0.17},  & \textrm{for }  h_b(2P)  \end{array}\right\}.
  % \]
\end{equation}

While one expected the decay to $h_b$ should be suppressed due to spin flip,
its higher rate was something puzzling.
The $\Upsilon(5S) \to h_b(nP)\pi^+ \pi^-$ decay mechanism seems to be exotic.
Belle found that
$\Upsilon(5S) \to Z_b^+ \pi^-$, then $Z_b^+$ decays to $h_b \pi^+$. $Z_b(10610)$
and $Z_b(10650)$ were found in $\Upsilon(1S)\pi^+\pi^-$,
$\Upsilon(2S)\pi^+\pi^-$, $\Upsilon(3S)\pi^+\pi^-$,
$h_b(1P)\pi^+\pi^-$, and $h_b(2P)\pi^+\pi^-$ decay with masses around the
$B^*B$ and $B^*B^*$ thresholds~\cite{Bondar}.
With more data, Belle II expects to
measure the mass and width more precisely. Further, Belle II can study neutral
$Z_b^0$  in $\Upsilon(5S) \to \Upsilon(nS) \pi^0\pi^0$~\cite{Krokovny}
and confirm in
other modes also.

Another study of interest to be done at Belle II is an energy scan.
A previous  energy scan of the $e+ e^- \to h_b(nP) \pi^+ \pi^-$ ($n=1,~2$)
cross sections by Belle gave first evidence for
$\Upsilon(6S) \to h_b(1P) \pi^+ \pi^-$
and observation of  $\Upsilon(6S) \to h_b(2P) \pi^+ \pi^-$. While studying the
resonant structure, they found evidence that it proceeds entirely via
the intermediate iso-vector states $Z_b(10610)$ and $Z_b(10650)$~\cite{Garmash}.
Currently only Belle II has the capability to do an $\Upsilon(nS)$ scan.

\begin{figure}[h!]
  \begin{center}
  \includegraphics[height=60mm,width=70mm]{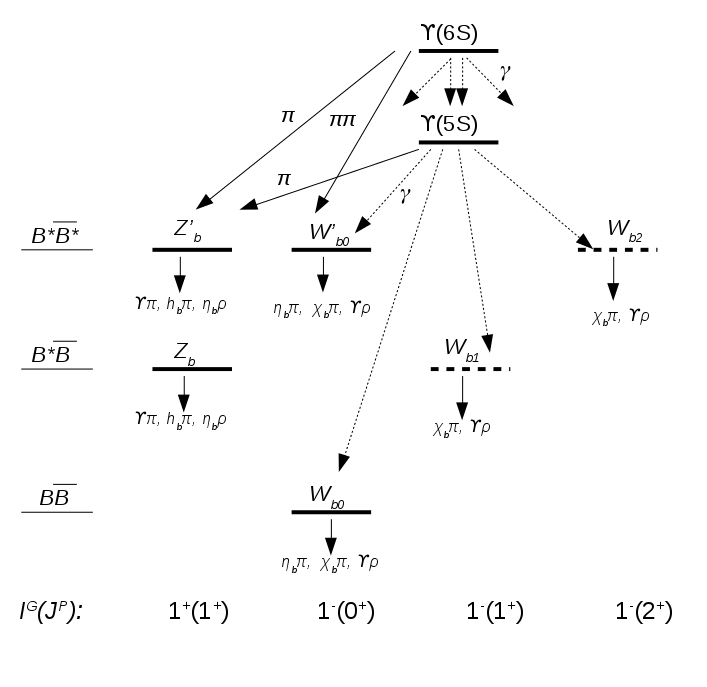}
  \caption{Accessing molecular partner bottomonium-like states ($W_b$) via
    transitions from $\Upsilon(5,6S)$.}
  \end{center}
  \label{fig:4}       % Give a unique label
\end{figure}

With a unique data set at $\Upsilon(6S)$, Belle II can study
$\Upsilon(6S) \to h_b(nP) \pi^+ \pi^-$,
$\Upsilon(6S) \to\Upsilon(mS) \pi^+ \pi^-$ $(n=1,~2;~m=1,~2,~3)$. If
$Z_b$ is a molecular state, then  Heavy Quark Spin symmetry suggests there should
be 2 or 4 molecular partner bottomonium-like states ($W_b$):
$\Upsilon(5S, 6S) \to W_{b0} \gamma$,  and
$\Upsilon(6S) \to W_{b0} \pi^+ \pi^-$, where
$W_{b0} \to \eta_b \pi, \to \chi_{b} \pi, \Upsilon \rho$. Fig.~\ref{fig:4}
summarizes the possible decays via which one can access the molecular
partners of bottomonium-like states~\cite{VoloshinPRF_84_031502_R_2011}

\subsection*{Acknowledgement}
I  would like to thank colleagues from the Belle II Collaboration. This
work is supported by the INSPIRE Faculty Award of the
Department of Science and Technology (India).


\end{document}